\documentclass{article}
\usepackage{spconf,amsmath,graphicx}

\usepackage{cite}
\usepackage{amsmath,amssymb,amsfonts}
\usepackage{graphicx}
\usepackage{subfigure}
\usepackage{textcomp}
\usepackage{multirow}
\usepackage{xcolor}
\usepackage{array}

\usepackage{algorithm,algpseudocode,float}
\usepackage[unicode=true,
 bookmarks=true,bookmarksnumbered=true,bookmarksopen=true,bookmarksopenlevel=1,
 breaklinks=false,pdfborder={0 0 0},pdfborderstyle={},backref=false,colorlinks=false]
 {hyperref}
\hypersetup{pdftitle={Your Title},
 pdfauthor={Your Name},
 pdfpagelayout=OneColumn, pdfnewwindow=true, pdfstartview=XYZ, plainpages=false}
\makeatletter
\newenvironment{breakablealgorithm}
{% \begin{breakablealgorithm}
	\begin{center}
		\refstepcounter{algorithm}% New algorithm
		\hrule height.8pt depth0pt \kern2pt% \@fs@pre for \@fs@ruled
		\renewcommand{\caption}[2][\relax]{% Make a new \caption
			{\raggedright\textbf{\ALG@name~\thealgorithm} ##2\par}%
			\ifx\relax##1\relax % #1 is \relax
			\addcontentsline{loa}{algorithm}{\protect\numberline{\thealgorithm}##2}%
			\else % #1 is not \relax
			\addcontentsline{loa}{algorithm}{\protect\numberline{\thealgorithm}##1}%
			\fi
			\kern2pt\hrule\kern2pt
		}
	}{% \end{breakablealgorithm}
		\kern2pt\hrule\relax% \@fs@post for \@fs@ruled
	\end{center}
}  

\title{Improving Self-supervised Learning for Out-of-distribution Task via Auxiliary Classifier}

\name{Harshita Boonlia$^{\star \dagger}$, Tanmoy Dam $^{\star}$ , Md Meftahul Ferdaus$^{\dagger}$, Sreenatha G. Anavatti$^{\star}$, Ankan Mullick$^{\star \dagger}$}

\address{$^{\star \dagger}$ CSE Department, Indian Institute of Technology Kharagpur \\
$^{\star}$ SEIT, University of New South Wales Canberra, Australia\\
$^{\dagger}$ ATMRI, Nanyang Technological University,
Singapore}

\DeclareUnicodeCharacter{2212}{-}    
    
\begin{document}

\maketitle

\begin{abstract}
In real world scenarios, out-of-distribution (OOD) datasets may have a large distributional shift from training datasets. This phenomena generally occurs when a trained classifier is deployed on varying dynamic environments, which causes a significant drop in performance. To tackle this issue, we are proposing an end-to-end deep multi-task network in this work. Observing a strong relationship between rotation prediction (self-supervised) accuracy and semantic classification accuracy on OOD tasks, we introduce an additional auxiliary classification head in our multi-task network along with semantic classification and rotation prediction head. To observe the influence of this addition classifier in improving the rotation prediction head, our proposed learning method is framed into bi-level optimisation problem where the upper-level is trained to update the parameters for semantic classification and rotation prediction head. In the lower-level optimisation, only the auxiliary classification head is updated through semantic classification head by fixing the parameters of the semantic classification head. The proposed method has been validated through three unseen OOD datasets where it exhibits a clear improvement in semantic classification accuracy than other two baseline methods. 
Our code is available at \url{https://github.com/harshita-555/OSSL}

\end{abstract}


%\end{keywords}
%

\section{Introduction}
In machine learning community, benchmarks like ImageNet \cite{deng2009imagenet}, CIFAR \cite{krizhevsky2009learning} etc. are commonly used to know the generalization ability of classifiers, where we assume that the test time input distributions are the same as the training distribution. However, when classifiers are applied to real-world applications like product recommendation, medical diagnosis, autonomous driving, they may face complex and dynamic shifts in the data distributions. Besides, new objects can be exposed to the classifiers at any time. Such issues in out-of-distribution (OOD) datasets may lead to catastrophic failure of the classifiers. In addition, annotations of test samples are not provided in many cases. Under such environment, classifiers' performance is usually evaluated by collecting new labeled test sets. Nonetheless, labeling adequate images in a novel scenario is very complex and highly expensive. To minimize such labeling cost, researchers have investigated various approaches for evaluating classifiers' performance on unlabeled test sets. Some researchers have developed complexity measurements on model parameters to analyse generalization of the classifiers \cite{arora2018stronger,corneanu2020computing,jiang2018predicting}. 

Researchers have proposed various methods to deal with OOD examples. For instance, probabilities from softmax distributions are utilized in \cite{hendrycks2016baseline} to detect wrongly classified and OOD examples. They have shown that OOD examples have a lower prediction probability than that of correct or in-sample examples. Researchers have also used self-supervision method \cite{deng2021does,sun2020test} to handle OOD tasks by introducing an auxiliary task that supports to create labels from unlabeled samples. In \cite{deng2021does}, they classified four different rotation angles of an image at $\{0^{\circ},~90^{\circ},~180^{\circ},~270^{\circ}\}$ to pay attention to the pretext task of rotation prediction. They jointly trained their network for both classification task and pretext task using CIFAR-10, MNIST, Tiny-ImageNet, and COCO, where they studied the correlation between those two task's accuracy. By considering many labeled testsets and plotting classification verses rotation prediction accuracy, a strong correlation (Pearson’s Correlation $r > 0.88$) is witnessed between those accuracies. Based on such finding, they learnt a linear regression model, which can predict classification accuracy on unseen test sets. They obtained ground truths from a given unlabeled test set by rotating images manually. It has been used to calculate the rotation prediction accuracy on the test images using the multi-task network. Afterward, this rotation prediction accuracy is used by the linear regression model to predict the semantic classification accuracy.

\begin{figure}
    \centering
    \includegraphics[scale=0.29]{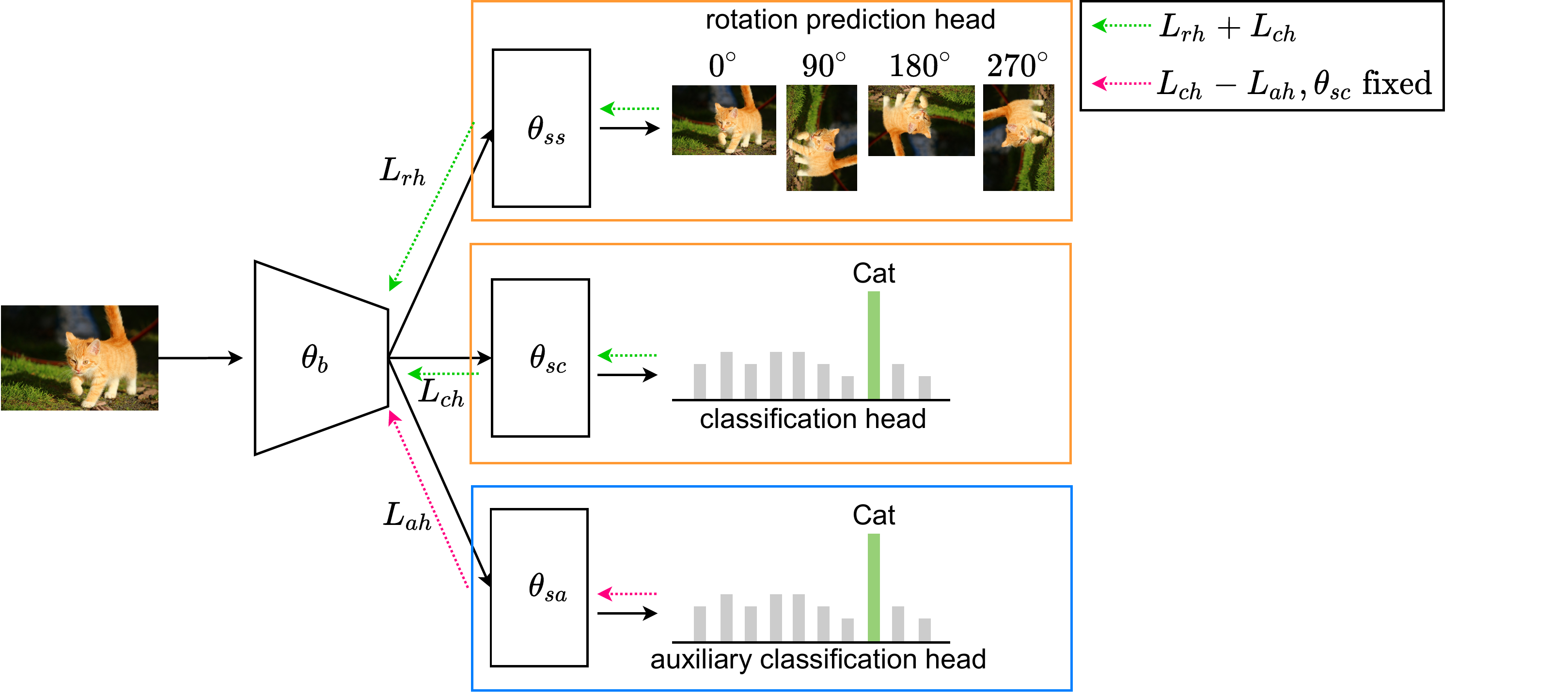}
    \caption{Proposed multi-task network structure for improving only the accuracy of rotation prediction via auxiliary classifier}
    \vspace{-2mm}
    \label{fig:proposed_architecture}
\end{figure}

Unlike the earlier work, in this work, we mainly focus on improving rotation prediction accuracy (self-supervised learning) using an auxiliary classifier for out of distribution task, which ultimately improves the semantic classification accuracy. Therefore, our proposed approach can be formulated as a bi-level optimization problem \cite{pham2020contextual}, where the rotation head and semantic classification head are learnt in the upper level. On the other hand, auxiliary classification head is learnt through the semantic classification head without updating the semantic classifier parameters in the lower level as shown in Fig. \ref{fig:proposed_architecture}. Since better rotation prediction accuracy indicates the model's higher ability to capture OOD features, we designed our multi-task architecture by focusing to maximize the performance of the rotation prediction head.             

Major contributions in this work can be stated as follows:

\begin{itemize}
  \item We propose a joint end-to-end multi-task framework called 'only self-supervised learning (OSSL)' for handling unseen OOD test sets. 
  \item We formulate the problem in a bi-level optimization fashion so as to improve semantic classification performance by maximizing the rotation prediction accuracy via an auxiliary classifier. 
  \item Our proposed framework has been validated using three unseen OOD data sets, where a better semantic classification accuracy have been witnessed in contrast with the baselines. 
\end{itemize}

\section{Proposed method}
In this paper, we have utilized a held-in training set and an unseen OOD test-set. We define the given training set as $D^{train}= \{(\textbf{x}_k, \textbf{y}_k)\}_{k=1}^N  \in (X \times Y)$ where, $x_k \in X $ is the $k$-th training image and $\textbf{y}_k =\{0, 1, ..., C-1\} \in Y$  corresponds to target variable spanning over $C$ classes. $N$ is the total number of samples present in the training dataset $D^{train}$. The OOD test set is defined in a similar fashion, $D^{test}=\{(\textbf{x}_l, \textbf{y}_l)\}_{l=1}^M$ where $\textbf{y}_l$ is target variable spanning over $C$ classes. Given a set of observation $(\textbf{x}_k, \textbf{y}_k) \subseteq (X, Y)$ drawn from a joint distribution $(\textbf{x}_k, \textbf{y}_k) \sim P_{XY}$, our objective is to design a robust classifier that can maximise classification accuracy for unseen OOD test-set.

%\subsection{Interrelated Study}
\subsection{Multi-task learning}
Researchers have observed in \cite{gidaris2018unsupervised} that a higher accuracy in rotation prediction indicates a model's superiority in capturing representation of learned features in lower dimensional manifold. Though they followed a completely unsupervised learning strategy, they did not consider the OOD task. In recent times, similar method has been developed to deal with OOD problem \cite{deng2021does}, where a linearly proportional relationship between rotation prediction and semantic classification has been observed. However, their proposed two-stage method can not provide an end-to-end solution. Besides, there are few questions that still remains in the end-to-end method. 
\begin{itemize}
    \item If rotational head estimates OOD classification accuracy perfectly, then how do we maximise the rotational head accuracy under multi-tasking learning framework? 
    \item How do we get end-to-end solutions for predicting OOD downstream task?
\end{itemize}  

% For doing that, they have assumed a set of synthetic dataset to maximise the rotational accuracy(self-supervised accuracy). Once the model is trained, model estimate testing labels from the unlabeled OOD test dataset. For rotational prediction head, they have used four rotational transformations to pass rotational head to calculate rotational accuracy. Afterwards, a linear relationship has been established through semantic accuracy and rotational prediction accuracy. Once the relationship has been established, they can estimates semantic accuracy for any given unlabeled dataset. However, they have not directly used test labels information for predicting test accuracy. It is not an end-to-end approach.  Besides, there are few questions still remains in the end-to-end method. 

% \begin{itemize}
%     \item If rotational head estimates OOD classification accuracy perfectly then how do I maximise the rotational head accuracy under multi-tasking learning framework? 
%     \item How do I get an end-to-end solutions for predicting OOD downstream task?
% \end{itemize}

\subsection{How to maximise the self-supervision (rotational-accuracy) task?}

\textbf{Network details:} To attain this objective, a multi-task learning framework has been developed for semantic classification along with self-supervised (rotation prediction) task. We utilise the multi-head network along with the same base network. Utilization of such multitasking framework is not enforcing  more complexity while improving the self-supervised task. To maximise the self-supervised performance on base-network, we introduce an auxiliary classifier along with semantic classifier head and self-supervised (rotation prediction) head. These minimal changes will not increase the burden on the base-network. For the base (feature extraction) network, we have taken a convolution neural network (e.g. densenet)followed by three fully connected layers for three different tasks. As depicted in Fig \ref{fig:proposed_architecture}, the base feature extractor is parameterised by $\theta_{b}$. All the remaining task-specific classification head are described as follows,  
\begin{itemize}
    \item semantic-classification prediction head is parameterised by $\theta_{sc}$.
    \item rotation prediction head is parameterised by $\theta_{ss}$.
    \item auxiliary semantic classification head is parameterised by $\theta_{sa}$.
\end{itemize}

\textbf{Rotational prediction head:}
We follow the similar rotation transformation as in \cite{deng2021does, gidaris2018unsupervised}. The four geometrical rotational transformations are applied to a train image($\textbf{x}$),   $F = \{G_{r}(\textbf{x}) \}$, where $G_{r}$ is the geometrical rotation function with four rotation angles $r = \{ 0^{\circ},  90^{\circ},180^{\circ},270^{\circ} \}$. This geometrical transformation can not alter the invariant nature \cite{dam2021improving}. Therefore, the rotational head can predict rotational accuracy by $4$-ways. 

% The rotational prediction defines us which rotation has been applied to image $(x)$; that will predict by $4$-way self-supervised task head. \\  

\textbf{Loss functions:}  The proposed OSSL method is associated with three individual classifications losses for three different tasks. The semantic classification loss is defined as follows,
\begin{align}
    L_{ch}= CE(\textbf{y}_{c}, \theta_{sc}(\theta_{b}(\textbf{x})))
\end{align}
where $CE=-\frac{1}{N}\sum_{y_{c}=1}^N \textbf{y}_{c}\log(\theta_{sc}(\theta_{b}(\textbf{x})))$

The rotation prediction classification loss is defined as follows,
\begin{align}
    L_{rh} = \frac{1}{4}\sum_{r \in \{ 0^{\circ},  90^{\circ},180^{\circ},270^{\circ}\} }CE(\textbf{y}_{r}, \theta_{ss}(\theta_{b}(G_{r}(\textbf{x}))))
\end{align}
where, $y_r$ is represented as one-hot-encode labels for all four rotational. 

The semantic auxiliary classification loss is defined as follows,
\begin{align}
    L_{ah}= CE(\textbf{y}_{c}, \theta_{sa}(\theta_{b}(\textbf{x})))
\end{align}
We have utilized the above three losses into a bi-level optimisation problem to maximise the self-supervision performance. In upper-level optimisation, the semantic classification head and rotation classification head parameters are learnt simultaneously to update the base-network parameters as well as corresponding task specific class parameters, where the objective can be expressed as follows:

\begin{equation}\label{eq:up_obj}
\underset{\theta_{b},\theta_{sc},\theta_{ss}}{\text{min}}L_{upper}
\end{equation}
where $L_{upper}=(L_{ch}+L_{rh})$. This upper level optimization problem is solvable with the stochastic gradient descent (SGD) method where it first tunes the parameters of both the task specific classifiers:
\begin{equation} \label{eq:upper_label}
    \{\theta_{sc},\theta_{ss}, \theta_{b} \}=\{\theta_{sc}, \theta_{ss}, \theta_{b} \}- l_r \sum_{ D^{train}}\nabla_{\{\theta_{sc},\theta_{ss}, \theta_{b}\}}L_{upper}
\end{equation}
where $l_r$ is the learning rate of the upper-level loop.

Similarly, in lower-level optimisation, the objective can be expressed as follows:

\begin{equation}\label{eq:low_obj}
\underset{\theta_{b},\theta_{sa}}{\text{min}}L_{lower}
\end{equation}
where $L_{lower}=(L_{ch}-L_{ah})$. As similar to upper level, the SGD method is used to optimize only the parameters of the auxiliary head and base network:
\begin{equation}\label{eq:lower_label}
    \{\theta_{sa}, \theta_{b} \}=\{\theta_{sa}, \theta_{b} \}- l_r \sum_{ D^{train}}\nabla_{\{\theta_{sa}, \theta_{b}\}}L_{lower}
\end{equation}
where, the same $l_r$ is being used to nullify the effects of semantic classification head in the backward path. However, the semantic classification head parameters ($\theta_{sc}$) remain fixed. For clarification, the proposed OSSL framework's learning strategy is given in Algorithm 1.

% The upper-level optimisation problem can follow the  To attain this objective, we first train the model with rotational head and classification head. For predicting the rotation head, we applied on rotational transformation to predict rotational accuracy. Afterwards, auxiliary classification head parameters are updated  to nullify the effects of semantic classification head in base-network. It is a two-stage update principle for maximising the rotational head prediction. Thus, we can formulate overall objective function as follows, 

% \begin{equation}
%     L_{tot} = L_{ch}+ L_{rh} + L_{ah}
% \end{equation}

% where, the three losses ($L_{ch}$,$L_{rh}$, $L_{ah}$) are  associated with three independent tasks respectively. $L_{ch}$ is related to the semantic classification head, $L_{rh}$ is the rotational prediction head and auxiliary classification head is $L_{ah}$. The learning strategy has been given in Algorithm. 

% The semantic classification is defined as follows,
% \begin{equation}
%     L_{ch} = \frac{1}{N}\sum_{y_k}y_{k}\log(\theta_{sc}(\theta_{b}(x_k)))
% \end{equation}

% The rotational prediction classification is defined as follows,
% \begin{equation}
%     L_{rh} =  \frac{1}{4N}\sum_{y_k \in \{ 0^{\circ},  90^{\circ},180^{\circ},270^{\circ}\} }y_{k}\log(\theta_{ss}(\theta_{b}(G_{r}(x_k)))
% \end{equation}
% The auxiliary semantic classification is defined as follows,
% \begin{equation}
%     L_{ah} = \frac{1}{N}\sum_{y_k}y_{k}\log(\theta_{sa}(\theta_{b}(x_k)))
% \end{equation}

%%%%%%%%%%%%%%%%%%%%%%%%%% %%%%%Algorithm 

\begin{breakablealgorithm}
    \caption{Learning Strategy of OSSL}
    \label{algorithm_ossl}
    \begin{algorithmic}[1]
        \State\textbf{Input}: training dataset $D^{train}$, testing dataset $D^{test}$, learning rates $l_r$, iteration numbers $n_{epoch}$
        \State\textbf{Output}: parameters of all the four networks $\{\theta_{b},\theta_{sc}, \theta_{ss}, \theta_{sa}\}$, 
        \For{$p=1$ to $n_{epoch}$}
        \State Update $\{\theta_{sc},\theta_{ss}, \theta_{b}\} $  parameters by using equation (\ref{eq:upper_label})   /* upper level optimisations/*
        \State Update $\{\theta_{sa}, \theta_{b}\} $  parameters by using equation (\ref{eq:lower_label}) when $\theta_{sc}$ is fixed   /* lower level optimisations/*
  %%/*only $L_{ch}$ is used but detach() the forward path/*
        \If{$p \geq 49 \, \& \, p \mathbin{\%}10 == 0$}
        \State calculate testing accuracy for $D^{test}$
        \EndIf
        \EndFor 
\end{algorithmic}
\end{breakablealgorithm}

\section{Experiments and Validation}
In this paper, our proposed OSSL method has been compared with two other baseline methods associated with two different losses, where the parameters of the first baseline is updated through semantic classification loss($L_{ch}$) and the second one is updated through semantic classification with rotational head losses ($L_{ch} + L_{rh}$). For experimental validations, two popular classification benchmark data sets have been considered to train the model such as: digits (MNIST) and natural image (CIFAR-10) data set. For both data sets, three different unseen OOD test-sets have been utilised to evaluate the model prediction accuracy. 

The LeNet-5 \cite{lecun2015lenet} model is a popular architecture for classifying the digit datasets (MNIST). Therefore, we consider LeNet-5 as a base feature extractor along with three classification heads. Original MNIST dataset is applied to train the model parameters, but, two different unseen OOD data sets namely USPS \cite{hull1994database} and SVHN \cite{netzer2011reading} are used to test it. Besides, both the unseen test-sets are having same number of classes($10$) as in the training set. Therefore, it is practical to use these data sets as  unseen OOD test-sets. On top of the backbone feature extractor i.e. LeNet-5, three tasks specific fully connected layers are being used. In addition, to analyse the effectiveness of the proposed method in a complex dataset, $Densenet-40$ ($40$ layers) architecture \cite{huang2017densely} is applied as a backbone feature extractor. In this case, CIFAR-10 is used to train the model, whereas $CIFAR-10.1$ is utilized as an unseen test-set to evaluate the model performance. CIFAR-10.1 is a popular benchmark dataset for the OOD classification task where collected test-set samples distributional shift cannot vary too much compared with the original CIFAR −10 samples \cite{recht2018cifar10.1, torralba2008tinyimages}. Moreover, CIFAR-$10.1$ dataset samples are subset of the Imagenet \cite{krizhevsky2012imagenet} dataset samples.

\begin{table}
\centering
\caption{The quantitative classification performance for unseen OOD test-set. While the base model is trained with held-in digit datasets, USPS and SVHN are being used as unseen(held-out) test sets. For CIFAR-10 train set, CIFAR 10.1 is used as unseen test-set. The reported classification accuracy is given in $\%$}
\begin{tabular}{c|c|c|c}

\hline 
Trainset & \multicolumn{2}{c|}{MNIST} & \multicolumn{1}{c}{CIFAR 10}\tabularnewline
\hline 
unseen OOD test-set & USPS & SVHN & CIFAR 10.1\tabularnewline
\hline 
$L_{ch}$ & 60.85 & \textbf{22.25} & 83.30\tabularnewline
\hline 
$L_{ch} + L_{rh}$ & 64.82 & 19.74 & 85.05\tabularnewline
\hline 
OSSL & \textbf{65.22} & 21.09 & \textbf{86.90}\tabularnewline
\hline 
\end{tabular}
\label{table: classification results}
\end{table}

Table \ref{table: classification results} represents the quantitative performance comparison for different unseen test-sets. For a fair comparison among these methods, the same base extractor is considered for all cases. For digit classification problem, original MNIST train-set is used to train the model. The obtained baseline classification performance (considering $L_{ch}$) for USPS testset is $60.85 \%$. A significant improvement in performance has been observed when rotation prediction loss is considered along with semantic classification loss. However, our proposed OSSL method has obtained best classification accuracy compared with the above two methods. The obtained classification accuracy is $65.22 \%$. However, for SVHN test-set, the best obtained accuracy is $22.25 \%$, which is from baseline when using only classification head. A significant performance drop is observed while considering the rational prediction head along with classification head. However, OSSL has obtained better model accuracy than the rotational head prediction, but it cannot outperform the baseline. Large distributional shift in the unseen SVHN test set samples compared to the held-in train set \cite{hsu2020generalized} could be the main reason for such performance deterioration. Such large distributional shift in dataset can't be represented by geometrical rotation with given held-in train-set. As a result, the classification performance is declined under multi-tasking learning framework. 

In addition, we have considered a complex CIFAR-10 dataset, where more complex network architecture has been utilised as a baseline. It is clearly observed from the Table \ref{table: classification results}, the proposed OSSL has outperformed the other two methods. While considering only classification head, the obtained accuracy is $83.30 \%$. A performance improvement in accuracy has been observed by considering both classification head and rotation prediction head, where the obtained classification accuracy is $85.05 \%$. The proposed OSSL method has attained best classification accuracy among all the three methods and the accuracy is $86.90 \%$. 

\begin{figure}
\begin{center}
\subfigure[][]{\includegraphics[scale=0.25]{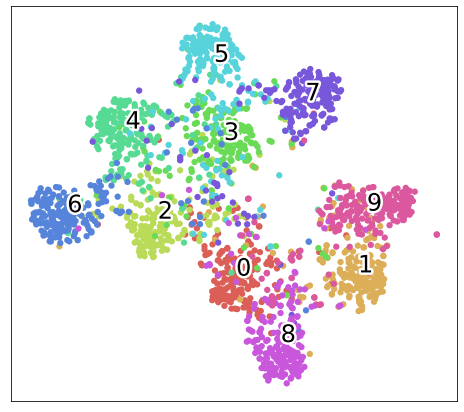}}
\subfigure[][]{\includegraphics[scale=0.24]{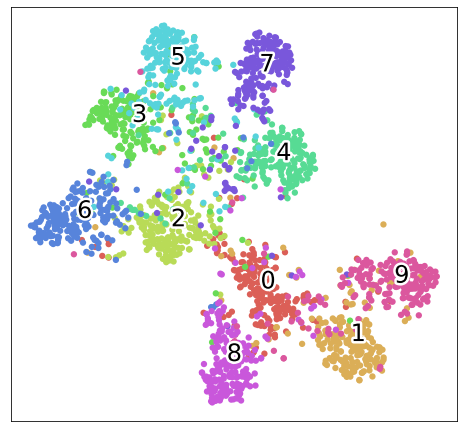}}
\subfigure[][]{\includegraphics[scale=0.25]{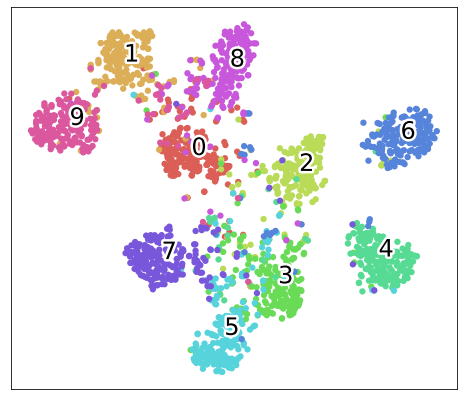}}
\caption{tSNE analysis on CIFAR-10.1 dataset using (a) baseline with $L_{ch}$, (b) baseline with $L_{ch}+ L_{rh}$, (c) proposed OSSL, where different classes are marked by digits (from 0 to 9)}
\label{fig:tsne}
\end{center}
\end{figure}

\subsection{tSNE analysis}
In this work, tSNE analysis \cite{van2008visualizing} is utilized to investigate the discriminative ability among different class distributions of our proposed methods and baselines as well. We considered test sets from CIFAR-10.1 dataset to project the original feature space to a two-dimensional space. In contrast to baselines, an effective separation among different classes is witnessed from OSSL as expected, where these classes are marked in different colors as shown in Fig. \ref{fig:tsne}. Such outcomes are also confirming that our proposed method can extract discriminative information from OOD datasets.

%From Fig. \ref{fig:tsne}, it is clearly observed that direct separation of the input data is not easy.

\section{Limitations}
The proposed OSSL method has focused on the maximisation of rotation prediction performance, which consequently improves the semantic classification accuracy. However, for some OOD datasets, incorporation of rotation prediction head along with a single semantic classification head can't guarantee to estimate a good performance on unseen test-set. As a consequence, rotation prediction head can influence the semantic classifier negatively and its performance can deteriorate. It may happen due to the sharing of common features by two classification heads. Moreover, the under-laying assumption is that rotation prediction head-based OOD task has to be well-defined and significant \cite{sun2020test}. Otherwise, rotational prediction head can't capture the significant features. For instance, from the SVHN dataset, a large distributional shift has been observed between training and testing dataset. In such case, though the proposed OSSL performs better than rotation prediction head and single semantic classifier-based method, it can not ensure its improvement over baseline.  %However, a classifier is called robust when it can detect and reject such large distributional shift from held-out dataset \cite{lee2018simple, hendrycks2016baseline, hsu2020generalized} .   

\section{Conclusion \& Future Works }
This paper presents a joint learning strategy to improve classification performance for  unseen OOD downstream task. For any OOD datasets, when a strong correlation is observed between rotation prediction accuracy and semantic classification accuracy, then we can maximise the rotation accuracy to obtain better semantic classification performance. To attain this objective, we formulated a bi-level optimisation framework where an additional auxiliary classifier was introduced to nullify the impact of semantic classification head  on base-feature extractors. The proposed OSSL method has been validated through three unseen OOD datasets. A significant improvement in classification performance is observed than the two other baselines. Some of the possible future directions utilizing our proposed learning strategy can be  stated as follows
\begin{itemize}
    \item To put a set of penalties for different dynamic test environments through invariant risk minimisation principle for handling large distributional shifts \cite{arjovsky2019invariant}.
    \item For handling class imbalance problem, a latent preserving GAN \cite{dam2021does, dam2021multi, dam2020mixture} can be used to generate minority class samples in dynamic tests environments.
    
    \item To handle adversarial robustness through maximising the rotation prediction accuracy \cite{chen2019improving}.
\end{itemize}

\section*{Acknowledgements}
T. Dam acknowledges UIPA funding from UNSW Canberra. 

\bibliographystyle{IEEEtran}
\bibliography{ref_ICIP22_hars}

\begin{thebibliography}{10}
\providecommand{\url}[1]{#1}
\csname url@samestyle\endcsname
\providecommand{\newblock}{\relax}
\providecommand{\bibinfo}[2]{#2}
\providecommand{\BIBentrySTDinterwordspacing}{\spaceskip=0pt\relax}
\providecommand{\BIBentryALTinterwordstretchfactor}{4}
\providecommand{\BIBentryALTinterwordspacing}{\spaceskip=\fontdimen2\font plus
\BIBentryALTinterwordstretchfactor\fontdimen3\font minus
  \fontdimen4\font\relax}
\providecommand{\BIBforeignlanguage}[2]{{%
\expandafter\ifx\csname l@#1\endcsname\relax
\typeout{** WARNING: IEEEtran.bst: No hyphenation pattern has been}%
\typeout{** loaded for the language `#1'. Using the pattern for}%
\typeout{** the default language instead.}%
\else
\language=\csname l@#1\endcsname
\fi
#2}}
\providecommand{\BIBdecl}{\relax}
\BIBdecl

\bibitem{deng2009imagenet}
J.~Deng, W.~Dong, R.~Socher, L.-J. Li, K.~Li, and L.~Fei-Fei, ``Imagenet: A
  large-scale hierarchical image database,'' in \emph{2009 IEEE conference on
  computer vision and pattern recognition}.\hskip 1em plus 0.5em minus
  0.4em\relax Ieee, 2009, pp. 248--255.

\bibitem{krizhevsky2009learning}
A.~Krizhevsky, G.~Hinton \emph{et~al.}, ``Learning multiple layers of features
  from tiny images,'' 2009.

\bibitem{arora2018stronger}
S.~Arora, R.~Ge, B.~Neyshabur, and Y.~Zhang, ``Stronger generalization bounds
  for deep nets via a compression approach,'' in \emph{International Conference
  on Machine Learning}.\hskip 1em plus 0.5em minus 0.4em\relax PMLR, 2018, pp.
  254--263.

\bibitem{corneanu2020computing}
C.~A. Corneanu, S.~Escalera, and A.~M. Martinez, ``Computing the testing error
  without a testing set,'' in \emph{Proceedings of the IEEE/CVF Conference on
  Computer Vision and Pattern Recognition}, 2020, pp. 2677--2685.

\bibitem{jiang2018predicting}
Y.~Jiang, D.~Krishnan, H.~Mobahi, and S.~Bengio, ``Predicting the
  generalization gap in deep networks with margin distributions,'' \emph{arXiv
  preprint arXiv:1810.00113}, 2018.

\bibitem{hendrycks2016baseline}
D.~Hendrycks and K.~Gimpel, ``A baseline for detecting misclassified and
  out-of-distribution examples in neural networks,'' \emph{arXiv preprint
  arXiv:1610.02136}, 2016.

\bibitem{deng2021does}
W.~Deng, S.~Gould, and L.~Zheng, ``What does rotation prediction tell us about
  classifier accuracy under varying testing environments?'' in
  \emph{International Conference on Machine Learning}.\hskip 1em plus 0.5em
  minus 0.4em\relax PMLR, 2021, pp. 2579--2589.

\bibitem{sun2020test}
Y.~Sun, X.~Wang, Z.~Liu, J.~Miller, A.~Efros, and M.~Hardt, ``Test-time
  training with self-supervision for generalization under distribution
  shifts,'' in \emph{International Conference on Machine Learning}.\hskip 1em
  plus 0.5em minus 0.4em\relax PMLR, 2020, pp. 9229--9248.

\bibitem{pham2020contextual}
Q.~Pham, C.~Liu, D.~Sahoo, and H.~Steven, ``Contextual transformation networks
  for online continual learning,'' in \emph{International Conference on
  Learning Representations}, 2020.

\bibitem{gidaris2018unsupervised}
S.~Gidaris, P.~Singh, and N.~Komodakis, ``Unsupervised representation learning
  by predicting image rotations,'' \emph{arXiv preprint arXiv:1803.07728},
  2018.

\bibitem{dam2021improving}
T.~Dam, S.~G. Anavatti, and H.~A. Abbass, ``Improving clustergan using
  self-augmented information maximization of disentangling latent spaces,''
  \emph{arXiv preprint arXiv:2107.12706}, 2021.

\bibitem{lecun2015lenet}
Y.~LeCun \emph{et~al.}, ``Lenet-5, convolutional neural networks,'' \emph{URL:
  http://yann. lecun. com/exdb/lenet}, vol.~20, no.~5, p.~14, 2015.

\bibitem{hull1994database}
J.~J. Hull, ``A database for handwritten text recognition research,''
  \emph{IEEE Transactions on pattern analysis and machine intelligence},
  vol.~16, no.~5, pp. 550--554, 1994.

\bibitem{netzer2011reading}
Y.~Netzer, T.~Wang, A.~Coates, A.~Bissacco, B.~Wu, and A.~Y. Ng, ``Reading
  digits in natural images with unsupervised feature learning,'' 2011.

\bibitem{huang2017densely}
G.~Huang, Z.~Liu, L.~Van Der~Maaten, and K.~Q. Weinberger, ``Densely connected
  convolutional networks,'' in \emph{Proceedings of the IEEE conference on
  computer vision and pattern recognition}, 2017, pp. 4700--4708.

\bibitem{recht2018cifar10.1}
B.~Recht, R.~Roelofs, L.~Schmidt, and V.~Shankar, ``Do cifar-10 classifiers
  generalize to cifar-10?'' 2018, \url{https://arxiv.org/abs/1806.00451}.

\bibitem{torralba2008tinyimages}
A.~Torralba, R.~Fergus, and W.~T. Freeman, ``80 million tiny images: A large
  data set for nonparametric object and scene recognition,'' \emph{IEEE
  Transactions on Pattern Analysis and Machine Intelligence}, vol.~30, no.~11,
  pp. 1958--1970, 2008.

\bibitem{krizhevsky2012imagenet}
A.~Krizhevsky, I.~Sutskever, and G.~E. Hinton, ``Imagenet classification with
  deep convolutional neural networks,'' \emph{Advances in neural information
  processing systems}, vol.~25, 2012.

\bibitem{hsu2020generalized}
Y.-C. Hsu, Y.~Shen, H.~Jin, and Z.~Kira, ``Generalized odin: Detecting
  out-of-distribution image without learning from out-of-distribution data,''
  in \emph{Proceedings of the IEEE/CVF Conference on Computer Vision and
  Pattern Recognition}, 2020, pp. 10\,951--10\,960.

\bibitem{van2008visualizing}
L.~Van~der Maaten and G.~Hinton, ``Visualizing data using t-sne.''
  \emph{Journal of machine learning research}, vol.~9, no.~11, 2008.

\bibitem{arjovsky2019invariant}
M.~Arjovsky, L.~Bottou, I.~Gulrajani, and D.~Lopez-Paz, ``Invariant risk
  minimization,'' \emph{arXiv preprint arXiv:1907.02893}, 2019.

\bibitem{dam2021does}
T.~Dam, M.~M. Ferdaus, S.~G. Anavatti, S.~Jayavelu, and H.~A. Abbass, ``Does
  adversarial oversampling help us?'' in \emph{Proceedings of the 30th ACM
  International Conference on Information \& Knowledge Management}, 2021, pp.
  2970--2973.

\bibitem{dam2021multi}
T.~Dam, N.~Swami, S.~G. Anavatti, and H.~A. Abbass, ``Multi-fake evolutionary
  generative adversarial networks for imbalance hyperspectral image
  classification,'' \emph{arXiv preprint arXiv:2111.04019}, 2021.

\bibitem{dam2020mixture}
T.~Dam, S.~G. Anavatti, and H.~A. Abbass, ``Mixture of spectral generative
  adversarial networks for imbalanced hyperspectral image classification,''
  \emph{IEEE Geoscience and Remote Sensing Letters}, 2020.

\bibitem{chen2019improving}
H.-Y. Chen, J.-H. Liang, S.-C. Chang, J.-Y. Pan, Y.-T. Chen, W.~Wei, and D.-C.
  Juan, ``Improving adversarial robustness via guided complement entropy,'' in
  \emph{Proceedings of the IEEE/CVF International Conference on Computer
  Vision}, 2019, pp. 4881--4889.

\end{thebibliography}

\end{document}